\documentclass[reprint, twocolumn, amsmath, amssymb, superscriptaddress]{revtex4-2}

\usepackage{graphicx}
\usepackage{dcolumn}
\usepackage{amsmath}

\usepackage[usenames,dvipsnames]{xcolor} 
\usepackage{hhline}
\usepackage{array}
\newcolumntype{P}[1]{>{\centering\arraybackslash}p{#1}}
\usepackage{textcomp}
\usepackage[normalem]{ulem}
\usepackage{changes}

\begin{document}

\title{Third-order exceptional point in an ion-cavity system}

\author{Jinuk Kim}
\thanks{Also at Department of Physics, Yale University, New Haven, Connecticut 06520, USA}
\thanks{These authors contributed equally to this work.}
\affiliation{Department of Electrical Engineering, Pohang University of Science and Technology (POSTECH), 37673 Pohang, Korea}
\author{Taegyu Ha}
\thanks{These authors contributed equally to this work.}
\affiliation{Department of Electrical Engineering, Pohang University of Science and Technology (POSTECH), 37673 Pohang, Korea}
\author{Donggeon Kim}
\affiliation{Department of Electrical Engineering, Pohang University of Science and Technology (POSTECH), 37673 Pohang, Korea}
\author{Dowon Lee}
\affiliation{Department of Electrical Engineering, Pohang University of Science and Technology (POSTECH), 37673 Pohang, Korea}
\author{Ki-Se Lee}
\affiliation{Department of Electrical Engineering, Pohang University of Science and Technology (POSTECH), 37673 Pohang, Korea}
\author{Jongcheol Won}
\affiliation{Department of Electrical Engineering, Pohang University of Science and Technology (POSTECH), 37673 Pohang, Korea}
\author{Youngil Moon}
\affiliation{Department of Electrical Engineering, Pohang University of Science and Technology (POSTECH), 37673 Pohang, Korea}
\author{Moonjoo Lee}
\email{moonjoo.lee@postech.ac.kr}
\affiliation{Department of Electrical Engineering, Pohang University of Science and Technology (POSTECH), 37673 Pohang, Korea}

\date{\today}

\begin{abstract}
We investigate a scheme for observing the third-order exceptional point (EP3) in an ion-cavity setting. 
In the lambda-type level configuration, the ion is driven by a pump field, and the resonator is probed with another weak laser field.
We exploit the highly asymmetric branching ratio of an ion's excited state to satisfy the weak-excitation limit, which allows us to construct  the non-Hermitian Hamiltonian $(H_{\textrm{nH}})$.
Via fitting the cavity-transmission spectrum, the eigenvalues of $H_{\textrm{nH}}$ are obtained.
The EP3 appears at a point where the Rabi frequency of the pump laser and the atom-cavity coupling constant balance the loss rates of the system. 
Feasible experimental parameters are provided.  
\end{abstract}

\maketitle

Exceptional point (EP) is a singularity in which two or more eigenvalues and eigenvectors coalesce in a parameter space~\cite{Wiersig2020}. 
In the vicinity of the EP, the non-Hermitian physical system exhibits unconventional phenomena that do not occur in the Hermitian system.
For example, strong spectral change was measured in photonics~\cite{Zhang2018a}, parity-time-symmetry-associated research~\cite{Rueter2010, Peng2016a} were performed~\cite{Feng2017}, and the topology was investigated extensively~\cite{Bergholtz2021}.
While such works were mostly done with the second-order EPs (EP2s), recent studies focused on multi-level structures where higher-order EPs can emerge~\cite{Mandal2021}.

More intriguing phenomena would be explored with the higher-order EPs.
The response of the system becomes more sensitive to external perturbations~\cite{Wu2021_2}, and the non-trivial topology would be more complex and richer as the order increases~\cite{Zhong2018, Patil2022}.
For instance, in the vicinity of EP, the response of the system to the perturbation $\epsilon$~$(\ll1)$ increases as $\epsilon^{1/N}$, where $N$ is the order of EP~\cite{Wu2021_2}.
Beyond encircling a point in a two-dimensional Riemann sheet~\cite{Doppler2016}, more complex topological structures were revealed with multiple EPs~\cite{Zhong2018} and also at higher dimensions~\cite{Patil2022}.
Robust topological state-switching based on EP3 was proposed~\cite{laha2020third}, an enhanced spontaneous emission at EP3 was predicted~\cite{lin2016enhanced}, and strong power amplification that scales with $N$ was investigated~\cite{Zhong2018a}.

Higher-order EPs have been observed in various experimental settings, including an acoustic cavity~\cite{Ding2016}, micro-ring~\cite{Hodaei2017}, dielectric sphere~\cite{Wang2019}, electronic circuit~\cite{Xiao2019}, acoustic metagrating~\cite{Fang2021}, and membrane~\cite{Patil2022}---these experiments were performed in classical domain.
Realization of EP in quantum realm~\cite{moiseyev2011non}, generally more difficult in practice~\cite{Wu2019}, would make quantum mechanics play in conjunction with this singularity: The EP effects would manifest by a post-selection measurement~\cite{Naghiloo2019}, and quantum noise can give an impact to the EP-based sensor~\cite{Lau2018}.
A recent work proposed a method for measuring the correlator in the Bose-Hubbard model using non-Hermitian perturbation~\cite{geier2022non}, and the spin transport was investigated in non-Hermitian quantum systems~\cite{lima2023spin}.

Here, we propose a scheme for observing an EP3 in a full quantum system, consisting of a single trapped ion coupled to an optical resonator. 
A lambda-type, atomic three-level configuration is considered. 
While a weak transition of the ion is pumped by a laser field from the cavity side, the other strong transition of the ion, to which the cavity frequency is tuned, is probed with a weak laser field along the cavity axis.
We exploit the highly asymmetric branching ratio of the ion's excited state to satisfy the weak-excitation limit: This allows us to neglect quantum jump operators and obtain $H_{\textrm{nH}}$.
By fitting the cavity-transmission spectrum, all three eigenvalues of $H_{\textrm{nH}}$ are obtained. 
We also identify that the EP3 appears at one point of the parameter space, where both the atom-cavity coupling constant $g$ and Rabi frequency $\Omega$ by the pump laser balance the atomic and cavity decay rates. 
From the perspective of atomic spectroscopy, the EP3 is equivalent to a crossover point from cavity electromagnetically induced transparency (EIT) to Autler-Townes splitting (ATS)~\cite{Anisimov2011, Zhu2013, Tan2014, Peng2014}.
Experimental parameters are examined with a $^{40}$Ca$^{+}$ ion and fiber-based cavity.

We remark that our system operates with trapped ions where the gate operation, qubit-state detection, and other controls can be done at a state-of-the-art level~\cite{Bruzewicz2019}.  
Here, we utilize one of the capabilities, nanoscopic position control~\cite{Guthoehrlein01}, to change one of the parameters, $g$, which constitutes a key ingredient for the observation of the EP3. 
Recently EP2s were observed with single trapped ions in free space~\cite{Wang2021_2,Ding2021}, and our work describes an extension of one more order with an optical resonator.
We also note that, while this ion-cavity settings have mostly been used for investigating quantum optics~\cite{Dubin10, Lee2019, Takahashi2020} and quantum networks,~\cite{Stute12, Walker2018, Krutyanskiy2023} to date, our work provides distinct research opportunities associated with non-Hermitian physics, employing both coherent ion-cavity interaction and cavity dissipation.

\begin{figure} [!t]
	\includegraphics[width=3.3in]{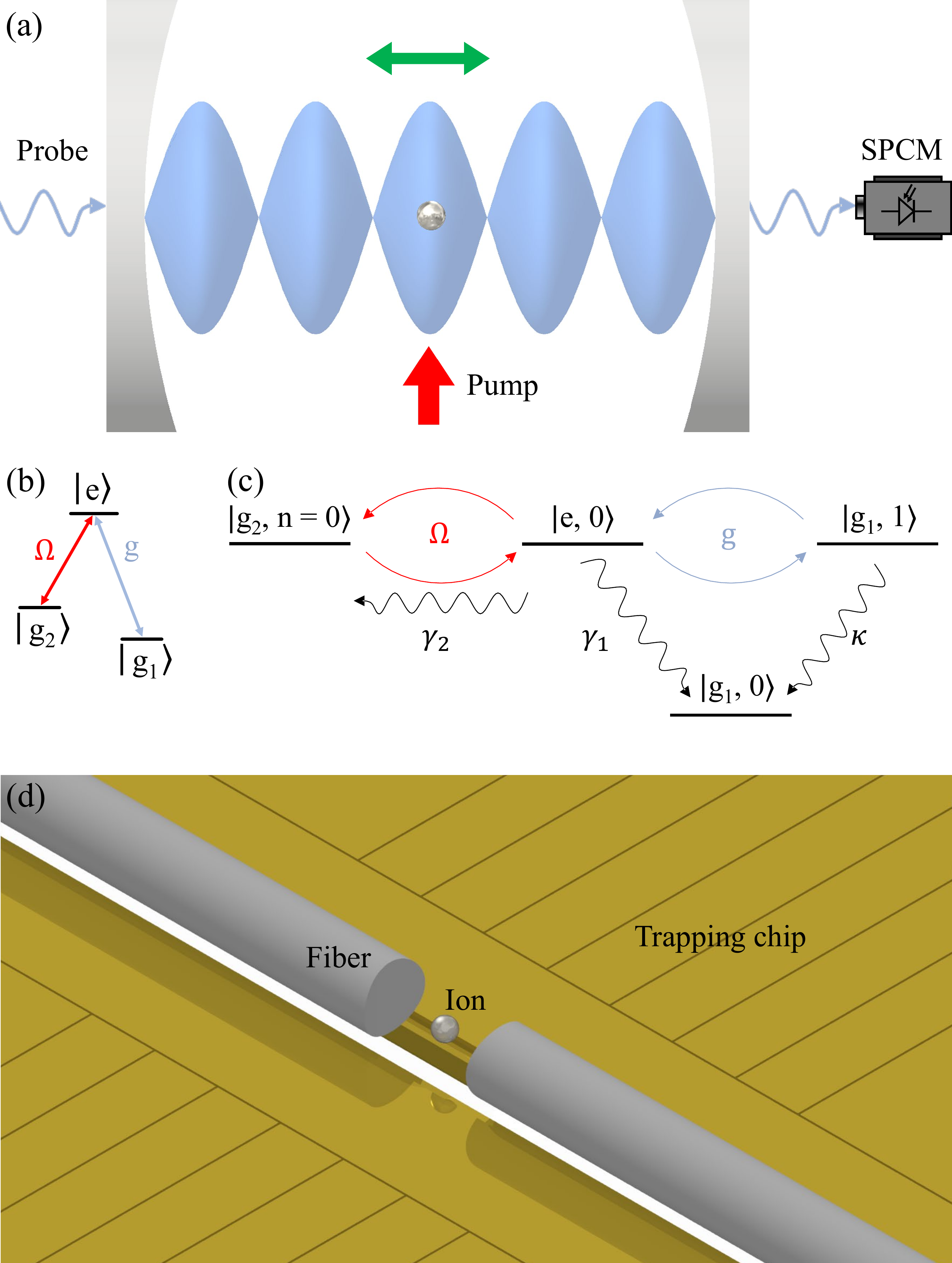} 
	\caption{
		(a) and (b) Trapped ion in an optical cavity.
		Laser from cavity side (red arrow) drives $|$g$_{2}\rangle$--$|$e$\rangle$ transition; $|$g$_{1}\rangle$--$|$e$\rangle$ transition is coupled to cavity. 
		Cavity is probed by a weak laser field and the output transmission is measured with a single photon counting module (SPCM).
		Ion position is movable along the cavity axis (green arrow). 
		(c) Energy levels of ion-cavity quantum states and their interactions. 
		$\Omega, \gamma_{1} \gg \gamma_{2}$.
		(d) Feasible experimental setting with the ion and fiber-based cavity. 
	}
	\label{fig:setting}
\end{figure}

We consider a setup in which a single trapped ion is coupled to an optical resonator (Fig.~\ref{fig:setting}).
The atomic energy levels consist of two ground states $|$g$_{1}\rangle$, $|$g$_{2}\rangle$ and one excited state $|$e$\rangle$.
The $|$g$_{2}\rangle$--$|$e$\rangle$ transition is pumped resonantly with a laser, from the cavity side (orthogonal to the cavity axis), of which we can change $\Omega$ by adjusting the laser power. 
The cavity frequency $\omega_{\textrm{c}}$ is resonantly tuned to the $|$g$_{1}\rangle$--$|$e$\rangle$ transition, and the cavity is weakly probed with a laser of the frequency $\omega_{\rm{p}}$. 
The ion position is movable along the cavity axis, by changing the voltages to the dc electrodes, for varying $g$~\cite{Guthoehrlein01}. 
This control can be done with a precision on the order of nanometers, so that we can change $g$ by moving the ion from a node to an antinode of the cavity field.
Consequently, via changing the laser power and the ion position, we steer an ion-cavity quantum state toward the EP3 in the parameter space of $g$ and $\Omega$.

The $H_{\textrm{nH}}$ for our ion-cavity state is given by

\begin{equation}
	H_{\rm_{nH}} = 
	\begin{pmatrix} 
		0 & \Omega/2 & 0 \\
		\Omega/2 & -i\gamma &  g \\
		0 & g & -i\kappa
	\end{pmatrix},
	\label{eq:nHH}
\end{equation}

\noindent
where the Planck constant $\hbar=1$, $\gamma  = \gamma_{1} +\gamma_{2}$ is the total atomic decay rate, $\gamma_{i} $ is the atomic decay rate from $|$e$\rangle$ to $|$g$_{i}\rangle$ $(i=1$ and $2)$, and $\kappa$ is the cavity decay rate. 
Three ion-photon states of $|$g$_2, n=0\rangle$, $|$e$, 0\rangle$, $|$g$_1, 1\rangle$ ($n$ indicates the number basis of cavity photons) comprise the bases of $H_{\rm{nH}}$.
We provide more details including the derivation of $H_{\rm{nH}}$ in the supplementary material.

\begin{figure} [!t]
	\includegraphics[width=3.3in]{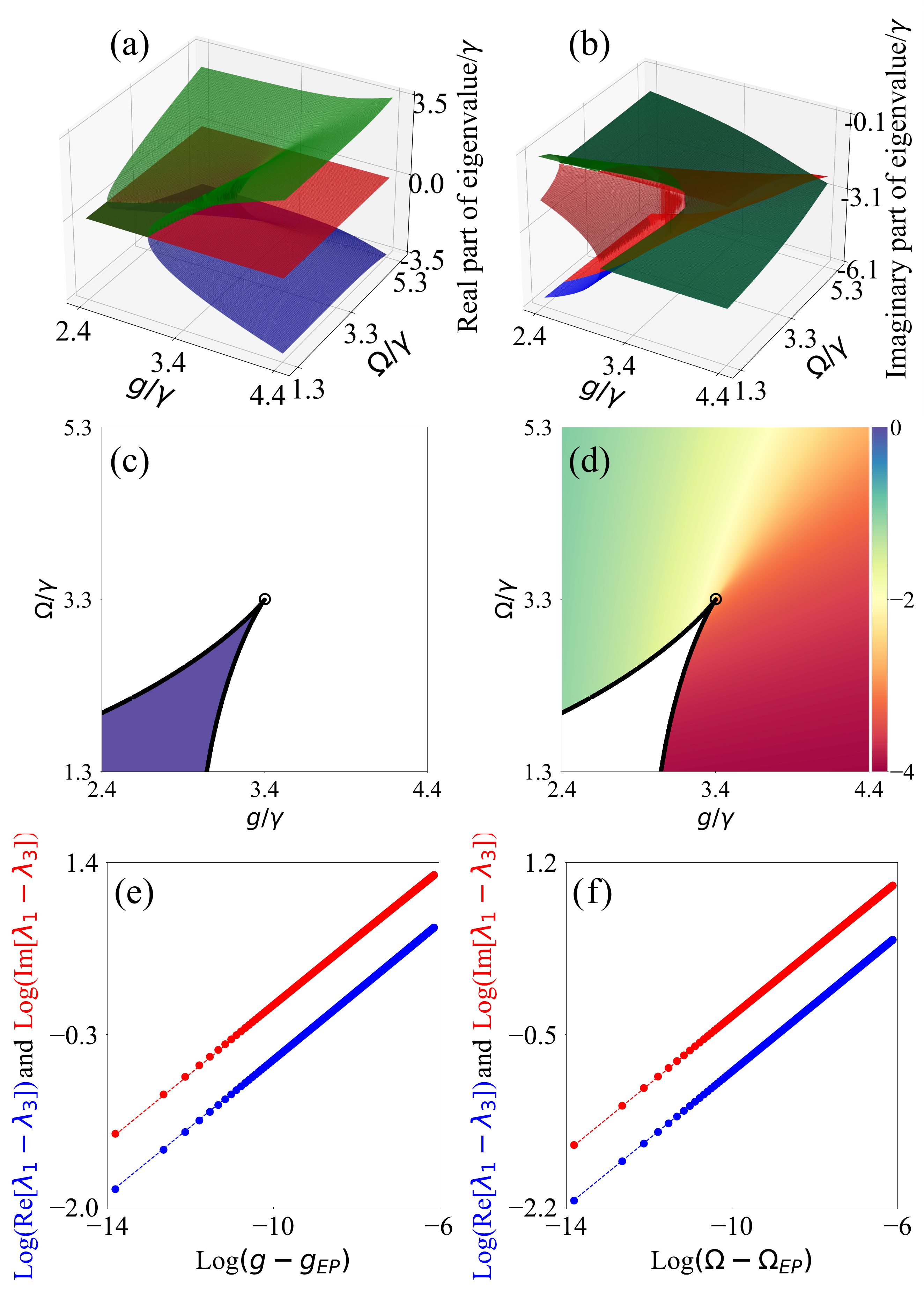} 
	\caption{
		(a) Real and (b) imaginary parts of eigenvalues as functions of $g$ and $\Omega$.
		Green, red, and blue sheets correspond to three eigensurfaces of $\lambda_{i}$~$(i=1, 2,3)$, respectively.
		$\kappa/\gamma=7.00$.
		(c) Slicecut of real-part eigensurfaces at Re[$\lambda_{i}$] = 0.
		Three sheets overlap at the purple region. 
		(d) Two surfaces of imaginary parts overlap at the colored region. 
		Black lines are EP2, and open black circles correspond to EP3.
		Log-log plot of real (blue circle) and imaginary (red circle) parts of eigenvalue differences $\lambda_{1}-\lambda_{3}$, and polynomial fits (lines) as a function of $g-g_{\rm{EP3}}$ in (e) and $\Omega-\Omega_{\rm{EP3}}$ in (f), in the vicinity of EP3.
	}
	\label{fig:eigensurface}
\end{figure}

\begin{figure*} [!t]
	\includegraphics[width=6.3in]{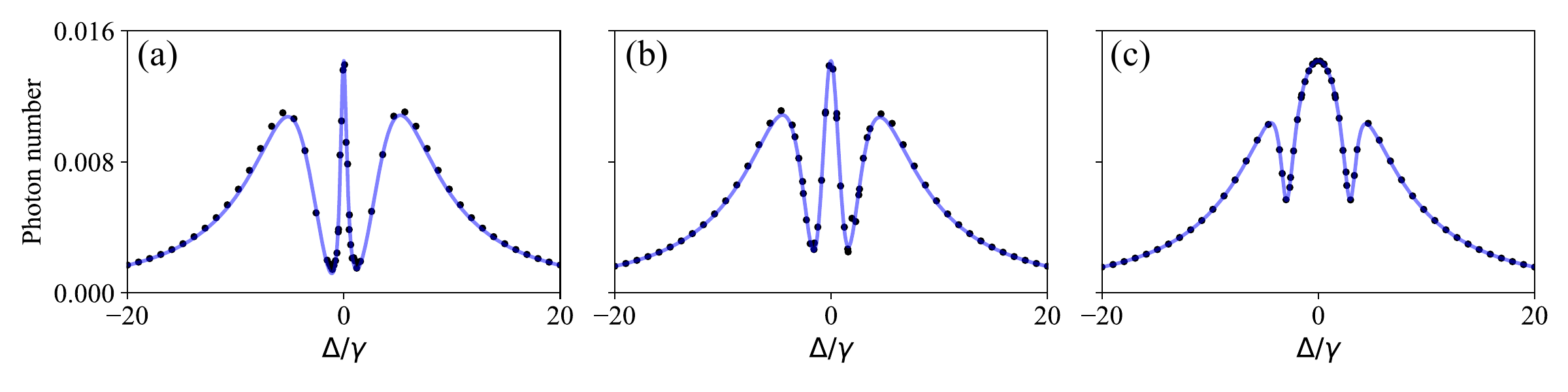} 
	\caption{
		Cavity photon number (cavity transmission) as a function of $\Delta(=\omega_{\textrm{p}}-\omega_{\textrm{c}})$.
		Black points are results of quantum Monte Carlo simulations, and the blue line is fitting result with Eq.~\eqref{eq:transmission}.
		Parameters are $g/\gamma=4.60$ and $\Omega/\gamma=2.00$ in (a), and $g_{\textrm{EP3}}/\gamma=3.41$ and $\Omega_{\textrm{EP3}}/\gamma=3.29$ in (b), and $g/\gamma=2.00$ and $\Omega/\gamma=6.00$ in (c).
		Spectrum of (b) is obtained at the EP3 condition.
		$\kappa/\gamma=7.00$. 
		Error bars (statistical errors) are smaller than symbol size.
	}
	\label{fig:transmission}
\end{figure*}

We discuss two conditions for obtaining the $H_{\rm{nH}}$. 
Note that, since the system needs to be operated in the weak-excitation limit for neglecting quantum jump operators, $|\textrm{g}_{1},0\rangle$ should be populated for most of the interaction time, i.e., $|\psi \rangle \simeq |$g$_{1},0\rangle + \alpha |$g$_{2},0\rangle + \beta |$e$, 0 \rangle + \delta |$g$_{1}, 1 \rangle$ ($|\alpha|, |\beta|, |\delta| \ll1 $).
The first condition consists of weak driving of the cavity, such that the cavity mean photon number is much less unity~\cite{Choi2010}: This would make the occupation of $|\textrm{g}_{1},1\rangle$ sufficiently small.
The second condition is that $\Omega$ and $\gamma_{1}$ must be much larger than $\gamma_{2}$, leading $|\textrm{e}, 0\rangle$ and $|\textrm{g}_{2}, 0\rangle$ to decay to $|\textrm{g}_{1}, 0\rangle$ immediately.
We can increase $\Omega$ by controlling the pump laser power, and satisfy  $\gamma_{1} \gg \gamma_{2}$ via choosing the ion species of $^{40}$Ca$^{+}$, which has an excited state $4^{2}$P$_{1/2}$ with very asymmetric decay rates.
While all simulations in the main text are done with $\gamma_{2}=0$, we compare this result to the simulations with actual values of $^{40}$Ca$^{+}$ ion, and find that the differences of the extracted eigenvalues are negligible.
More details are described in the supplementary material.

Next, we obtain the eigenvalues of $H_{\rm{nH}}$ as functions of $g$ and $\Omega$ numerically. 
Figures~\ref{fig:eigensurface}(a) and~\ref{fig:eigensurface}(b) show three Riemann sheets with real and imaginary parts of the eigenvalues, respectively. 
The EP3 appears at the position where the three eigensurfaces coalesce, with $\lambda_{\rm{EP3}}=-i(\gamma + \kappa)/3$ at $g_{\rm{EP3}}$ and $\Omega_{\rm{EP3}}$.
We denote $g_{\rm{EP3}}$ and $\Omega_{\rm{EP3}}$ as the atom-cavity coupling constant and the Rabi frequency of the pump laser at the EP3, respectively.
We also investigate the scaling behavior of the eigenvalues in the vicinity of the EP3 [Figs.~\ref{fig:eigensurface}(e) and~\ref{fig:eigensurface}(f)].
The real and imaginary parts, Log$[\textrm{Re}[\lambda_{1}-\lambda_{3}]]$ and Log$[\textrm{Im}[\lambda_{1}-\lambda_{3}]]$ are fit with $a + \frac{1}{3}\cdot\textrm{Log} \left( g-g_{\rm{EP3}} \right) $ and $a'+\frac{1}{3} \cdot \textrm{Log} \left( \Omega-\Omega_{\rm{EP3}} \right)$ with the coefficients $a$ and $a'$, showing good agreement with the calculation results.
Calculated $R^{2}$ values are larger than 0.99 for all data in Figs.~\ref{fig:eigensurface}(e) and~\ref{fig:eigensurface}(f).
This agreement presents that our singular point exhibits a representative feature of the EP3.

We point out that the manipulation of cavity and laser frequencies would allow us to envisage a parameter space of higher dimensions. 
In Fig.~\ref{fig:eigensurface}, our control parameters are $g$ and $\Omega$, which present similar topological structures with Refs.~\cite{laha2020third,lin2016enhanced,zhang2019higher}.
Including the frequency differences between the cavity and $|\textrm{g}_{1}\rangle$--$|\textrm{e}\rangle$ transition and the pump laser and $|\textrm{g}_{2}\rangle$--$|\textrm{e}\rangle$ transition, the Riemann sheets appear in three or four dimensional parameter spaces.
This would offer opportunities for studying more complex topology, such as like exceptional lines~\cite{Pap2018}, knots~\cite{Patil2022}, and exceptional surfaces~\cite{Wang2020b}.

\begin{figure*} [t]
	\includegraphics[width=6.3in]{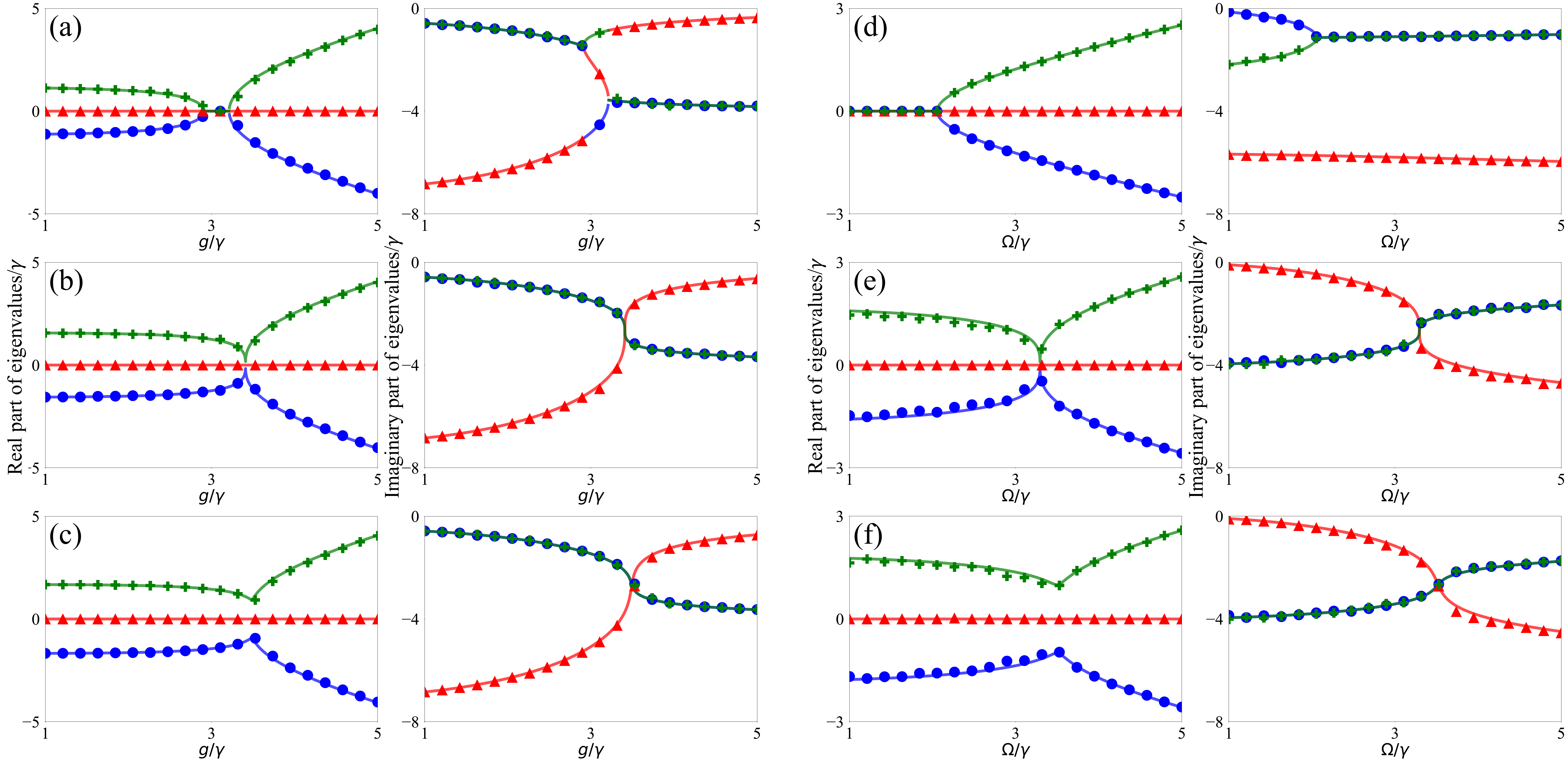} 
	\caption{
		Real and imaginary parts of the numerically obtained eigenvalues of $H_{\rm{nH}}$ are shown in lines.
		Green line is $\lambda_{1}$, red is $\lambda_{2}$, and blue is $\lambda_{3}$.
		Points are obtained from fitting the transmission spectrum, results of quantum Monte Carlo simulation, with Eq.~\eqref{eq:transmission}.  
		(a) $\Omega/\gamma = 2.50$, (b) $\Omega/\gamma= 3.29$, (c) $\Omega/\gamma=3.50$, (d) $g/\gamma = 2.50$, (e) $g/\gamma= 3.41$ and (f) $g/\gamma=3.50$. 
		EP3 appears in (b) and (e). 
	}
	\label{fig:comparison}
\end{figure*}

We proceed to describe our scheme for extracting eigenvalues from actual experiments.
We ``simulate'' the experiment with a quantum Monte Carlo (qMC) calculation that includes the effect of statistical errors of real experiments.
Our calculation consists of time evolution by non-Hermitian effective Hamiltonian with randomly occurring quantum jump operators~\cite{Carmichael2008}; more details are given in supplementary material.
Each data point in Fig.~\ref{fig:transmission} corresponds to averaging about $10,000$ simulation results in which the time duration of each trajectory is $\sim 50/\gamma$.
Such repetition and measurement duration are realistic in the ion-cavity experiments.
The cavity is probed at the bare cavity mean photon number 0.014 such that the weak-excitation condition is satisfied.
For three sets of $g$ and $\Omega$ the simulation results are presented in Fig.~\ref{fig:transmission}.
When $g>\Omega$, cavity EIT gives rise to the central transparency peak with vacuum-Rabi spectrum [Fig.~\ref{fig:transmission}(a)]; when $g<\Omega$, in a broad cavity spectrum, we point out two ``dips'' that are caused by the pump-induced dressed states, which reflects the pronounced feature of ATS [Fig.~\ref{fig:transmission}(c)]; the EP3 appears in-between [Fig.~\ref{fig:transmission}(b)].

These spectral features are also manifested in the extracted eigenvalues of Fig.~\ref{fig:comparison} (see below). 
Similarly with Fig.~\ref{fig:transmission}(a), we find the cavity-EIT region, for instance, $g/\gamma=5.00$, in Fig.~\ref{fig:comparison}(a).
The center frequencies of three peaks appear in real parts, and the imaginary parts are associated with widths, with a notably narrow spectrum of the dark state (red triangle of imaginary part). 
On the other regime of ATS, e.g.~$\Omega/\gamma=5.00$ in Fig.~\ref{fig:comparison}(d), the broad cavity-photon state contributes to a large imaginary part close to the width of the bare cavity (red triangle), with those of the dressed states (blue circle and green cross).
In the middle of the two regions, the EP3 emerges at a crossover point where all eigenvalues coalesce.

We then extract the eigenvalues $\lambda_{i}$~$(i = 1, 2, 3)$ of $H_{\rm{nH}}$, in a given $g$ and $\Omega$, from the qMC simulated spectrum. 
We formulate the equation of motion with $H_{\rm{nH}}$ as

\begin{equation}
	i\frac{d}{dt} |\psi \rangle = \left( H_{\rm{nH}} - \Delta I \right) |\psi\rangle + 
	\begin{pmatrix}
		0  \\
		0  \\
		\epsilon \\
	\end{pmatrix},	
	\quad
	|\psi\rangle = 
	\begin{pmatrix}
		\alpha  \\
		\beta  \\
		\delta \\
	\end{pmatrix}
\end{equation}

\noindent
where $\Delta=\omega_{\rm{p}}-\omega_{\rm{c}}$, and $\epsilon$ denotes the weak amplitude of the probe laser.
Correspondingly, we find an analytic expression of the cavity transmission  

\begin{equation}
	T = \left | \frac{\Delta (  i\kappa -\Delta + \lambda_{2} + \lambda_{3} ) \kappa + \lambda_{1} (\Delta \kappa - i \lambda_{2} \lambda_{3} ) }{(\Delta - \lambda_{1})(\Delta - \lambda_{2}) (\Delta - \lambda_{3} ) } \right |^2,
	\label{eq:transmission}
\end{equation}

\noindent
and through fitting the measured spectrum with Eq.~\eqref{eq:transmission}, three eigenvalues of $\lambda_{i}$ are obtained. 
More details are described in the supplementary material.
The fitting results are shown in Fig.~\ref{fig:transmission}, agreeing with the qMC simulation spectrum very well, quantified with $R^{2}$ values larger than 0.99 for all results of Figs.~\ref{fig:transmission}(a)-\ref{fig:transmission}(c).

Figure~\ref{fig:comparison} shows the extraction of the eigenvalues with this model for several $g$ and $\Omega$. 
The solid lines are real and imaginary parts of the numerically obtained eigenvalues of $H_{\rm{nH}}$.
The points correspond to real and imaginary parts, extracted from fitting the transmission spectrum with Eq.~\eqref{eq:transmission}.
Numerically obtained eigenvalues agree well with the extraction from the fitting results, also confirming that our model is correct.

Finally, we seek a feasible experimental setting to observe the EP3. 
We consider that a $^{40}$Ca$^{+}$ ion~\cite{Stute12, Lee2019, Takahashi2020} is coupled to a fiber-based optical resonator.
We choose the strong 4$^{2}$S$_{1/2}$ to 4$^{2}$P$_{1/2}$ transition [$\gamma_{1}/(2\pi) = 10.8$~MHz] at $397$~nm, coupled to the cavity.
The specification of the cavity includes radii of curvature of 350~$\mu$m for both fiber mirrors, a cavity length of 500~$\mu$m, a maximum atom-cavity coupling constant $g_{0}/(2\pi)=49.7$~MHz, and $\kappa/{(2\pi)}=80.0$~MHz (finesse of $\sim 1880$)~\cite{Ballance2017, Lee2019a, Ong2020}.
The 3$^{2}$D$_{3/2}$ to 4$^{2}$P$_{1/2}$ transition [$\gamma_{2}/(2\pi)=0.74$~MHz] at 866~nm corresponds to the weak transition, pumped by a power-tunable laser from the cavity side.
In this configuration, the 4$^{2}$S$_{1/2}$ state is mostly occupied for satisfying the weak-excitation condition.
We expect that $g_{\rm{EP3}}/(2\pi)=38.9$~MHz, which would be reached by moving the ion position between an antinode and a node of the cavity field; $\Omega_{\textrm{EP3}}/(2\pi)=37.7$~MHz. 
Asymmetric mirror coating would make the transmission measurement more effective~\cite{Lee2022}.

In conclusion, we have theoretically explored the emergence of EP3 and a measurement method in an ion-cavity system. 
In the lambda-type level configuration, the EP3 appears in the parameter space of $g$ and $\Omega$.
The eigenvalues of $H_{\rm{nH}}$ are obtained through fitting the cavity-transmission spectrum.
Our work provides interesting possibilities to investigate the non-Hermitian physics in the ion-cavity setting.

The supplementary material consists of five sections. 
The first part consists of the derivation of $H_{\rm{nH}}$. 
The second one includes eigenvalues and eigenvectors of $H_{\rm{nH}}$ at the EP3. 
Quantum Monte Carlo simulation is described in the third section, followed by the next part addressing the derivation of Eq.~\eqref{eq:transmission}.
The material closes with the simulations with actual values of a $^{40}$Ca$^{+}$ ion.

We thank Kyungwon An for insightful comments.
This work has been supported by BK21 FOUR program and Educational Institute for Intelligent Information Integration, the National Research Foundation (Grant Nos.~2019R1A5A1027055 and~2020R1I1A2066622), the Institute for Information \& communications Technology Planning \& Evaluation (IITP, Grant No.~2022-0-01040), the Samsung Science and Technology Foundation (Grant Nos.~SSTF-BA2101-07 and~SRFC-TC2103-01), and the Samsung Electronics Co., Ltd. (Grant No.~IO201211-08121-01).

The data that support the findings of this study are openly available at https://doi.org/10.5281/zenodo.7821387.

\bibliographystyle{apsrev4-2}
\bibliography{bibliography}

\clearpage
\onecolumngrid
\appendix

\section{non-Hermitian Hamiltonian}

We derive the non-Hermitian Hamiltonian, Eq.~(1) of the main text.
We begin with the total Hamiltonian describing the system of the ion, cavity field, and pump laser:

\begin{equation}
	H_{\textrm{tot}} = \omega_{1} \sigma_{\textrm{e}} + (\omega_{1}-\omega_{2}) \sigma_{\textrm{g}_{2}} +  \omega_{\textrm{c}} a^{\dagger}a
	+ g ( a^{\dagger}\sigma_{\textrm{g}_{1}\textrm{e}} + \textrm{h.~c.} )
	+  \frac{\Omega}{2} ( \sigma_{\textrm{g}_{2}\textrm{e}} e^{i\omega_{\textrm{p}\rm{'}}t} + \textrm{h.~c.} ),
	\label{eq:Htot}
\end{equation}

\noindent
where the Planck constant $\hbar=1$, $\omega_{i}$ is the transition frequency between $|$g$_{i}\rangle$ and $|$e$\rangle$ $(i=1, 2)$, $\sigma_{\textrm{e(g}_{i})}$ is the projection operator onto the state $|\textrm{e}\rangle$($|\textrm{g}_{i}\rangle$), $\omega_{\textrm{c}}$ is the cavity frequency, $\omega_{\rm{p'}}$ is the frequency of the pump laser, $a^{\dagger}(a)$ is the cavity photon creation (annihilation) operator, $g$ is the atom-cavity coupling constant, $\sigma_{\textrm{g}_{i}\textrm{e}}=\sigma_{\textrm{eg}_{i}}^{\dagger}$ is the transition operator between $|$e$\rangle$ and $|$g$_{i}\rangle$, and $\Omega$ is the Rabi frequency by the pump laser.
The interaction Hamiltonian is obtained by using the relation $H_{\rm{int}}=i \dot{U}U^{\dagger} + UH_{\rm{tot}}U^{\dagger}$ where $U=\textrm{exp}\left[ i \left( \omega_{1}\sigma_{\textrm{e}} + (\omega_{1}-\omega_{\textrm{p}\rm{'}} )\sigma_{\textrm{g}_{2}} + \omega_{\textrm{c}} a^{\dagger}a    \right)t \right] $,

\begin{equation}
	H_{\textrm{int}} = \Delta_{\textrm{p}\rm{'}} \sigma_{ \textrm{g}_{2} } 
	+ g \left( a^{\dagger} \sigma_{\rm{g}_{1}\rm{e}} + \textrm{h.~c.} \right)
	+ \frac{\Omega}{2} \left( \sigma_{\textrm{g}_{2}\textrm{e}} + \textrm{h.~c.} \right),
\end{equation}

\noindent
with $\Delta_{\textrm{p}\rm{'}} = \omega_{\textrm{p}\rm{'}} - \omega_{2} =0$ and $\omega_{1}=\omega_{\textrm{c}}$ in our work. 

We proceed to include the dissipation operators in the Lindblad form. 
The equation of motion for the ion-cavity density matrix $\rho$ is expressed as 

\begin{equation}
	\frac{d\rho}{dt}  = -i \left[ H_{\textrm{int}}, \rho \right] 
	+ \sum_{i=1,2} {\gamma_{i} \left( 2 \sigma_{\textrm{g}_{i}\textrm{e}} \rho \sigma_{\textrm{eg}_{i}} 
		- \sigma_{\textrm{eg}_{i}} \sigma_{\textrm{g}_{i}\rm{e}} \rho - \rho \sigma_{\textrm{eg}_{i}} \sigma_{\textrm{g}_{i}\textrm{e}}  \right)}
	+ \kappa \left( 2a \rho a^{\dagger} - a^{\dagger}a\rho - \rho a^{\dagger}a \right).
	\label{eq:eom}
\end{equation}

\noindent
where $\gamma_{i}$ is the atomic decay rate from $|\textrm{e}\rangle$ to $|\textrm{g}_{i}\rangle$, and $\kappa$ is the cavity decay rate.

Given Eq.~\eqref{eq:eom}, the terms are rearranged for obtaining

\begin{align}
	\label{eq:eom_re}
	\frac{d\rho}{dt}  = & -i\left[ g \left( a^{\dagger} \sigma_{\textrm{g}_{1}\textrm{e}} + \textrm{h.~c.} \right) 
	+ \frac{\Omega}{2}  \left( \sigma_{\textrm{g}_{2}\textrm{e}} + \textrm{h.~c.} \right), \rho \right] \nonumber		           
	- i \sum_{i=1,2} \gamma_{i} \left[ -i \sigma_{\textrm{eg}_{i}}  \sigma_{\textrm{g}_{i}\textrm{e}} , \rho \right]
	- i \kappa \left[ -i a^{\dagger}a, \rho \right] \\ 				       
	& 
	+2 \sum_{i=1,2} { \gamma_{i}  \sigma_{\textrm{g}_{i}\rm{e}}  \rho  \sigma_{\textrm{eg}_{i}}  }
	+ 2 \kappa a\rho a^{\dagger}  \nonumber \\ \nonumber
	\simeq &  -i \left[   g \left( a^{\dagger} \sigma_{\textrm{g}_{1}\textrm{e}} + \textrm{h.~c.} \right) 
	+ \frac{\Omega}{2} \left( \sigma_{\textrm{g}_{2}\rm{e}} + \textrm{h.~c.} \right)	
	+ \sum_{i=1,2}  -i\gamma_{i} \sigma_{\textrm{e}}  
	- i \kappa a^{\dagger}a 
	, \rho  \right] \\ 
	= & -i \left[ H_{\textrm{nH}}', \rho \right].                             
\end{align}

Since our system operates in the weak-excitation limit, the state is mostly populated in the ground state $| \textrm{g}_{1}, 0 \rangle$, and thus the quantum jump operators (the second line of Eq.~\eqref{eq:eom_re}) are neglected~\cite{Carmichael2008}. 
Then we find the equation of motion with the non-Hermitian Hamiltonian $H_{\textrm{nH}}'$.


Now we obtain $H_{\rm{nH}}$ in a $3 \times 3$ matrix form. 
We define the four involved states as the following basis

\begin{equation}
	|\textrm{g}_{1}, n=0 \rangle =
	\begin{pmatrix}
		1 \\
		0 \\
		0 \\
		0
	\end{pmatrix}
	\quad
	|\textrm{g}_{2}, 0 \rangle =
	\begin{pmatrix}
		0 \\
		1 \\
		0 \\
		0
	\end{pmatrix}
	\quad
	|\textrm{e}, 0 \rangle =
	\begin{pmatrix}
		0 \\
		0 \\
		1 \\
		0
	\end{pmatrix},
	\quad	
	|\textrm{g}_{1}, 1 \rangle =
	\begin{pmatrix}
		0 \\
		0 \\
		0 \\
		1
	\end{pmatrix}		
\end{equation}

\noindent
where $n$ is the basis of cavity-photon numbers.
The operators are represented by

\begin{equation}
	a =
	\begin{pmatrix}
		0 & 0 & 0 & 1 \\
		0 & 0 & 0 & 0 \\
		0 & 0 & 0 & 0  \\
		0 & 0 & 0 & 0 
	\end{pmatrix}
	\quad
	\sigma_{\rm{g}_{2}\rm{e}} =
	\begin{pmatrix}
		0 & 0 & 0 & 0 \\
		0 & 0 & 1 & 0 \\
		0 & 0 & 0 & 0  \\
		0 & 0 & 0 & 0 
	\end{pmatrix}
	\quad
	\sigma_{\rm{g}_{1}\rm{e}} =
	\begin{pmatrix}
		0 & 0 & 1 & 0 \\
		0 & 0 & 0 & 0 \\
		0 & 0 & 0 & 0  \\
		0 & 0 & 0 & 0 
	\end{pmatrix}.
\end{equation}

With these bases and operators, it is straightforward to show that $H_{\rm{nH}}'$ is given by

\begin{equation}
	\label{eq:H_nh'}
	H'_{\rm{nH}} =  
	\begin{pmatrix}
		0 & 0 & 0 & 0 \\
		0 & 0 & \Omega / 2 & 0 \\
		0 & \Omega / 2 & -i\gamma & g  \\
		0 & 0 & g & -i\kappa 
	\end{pmatrix},
\end{equation}

\noindent
with $\gamma = \gamma_{1} + \gamma_{2}$ $(\gamma_{1}\gg\gamma_{2})$. 
The $3 \times 3$ sub-matrix, excluding the first row and column of $H'_{\rm{nH}}$, corresponds to $H_{\rm{nH}}$ of the main text:

\begin{equation}
	\label{eq:H_nh}
	H_{\rm{nH}} = 
	\begin{pmatrix} 
		0 & \Omega/2 & 0 \\
		\Omega/2 & -i\gamma &  g \\
		0 & g & -i\kappa
	\end{pmatrix}.
\end{equation}

\section{Eigenvalue and eigenvector of $H_{\textrm{nH}}$ at the third-order exceptional point}

\subsection{Eigenvalue}
We describe an analytic form of the eigenvalue of $H_{\rm{nH}}$ at the third-order exceptional point (EP3).
We also find the Rabi frequency $\Omega_{\textrm{EP3}}$ and the atom-cavity coupling constant $g_{\textrm{EP3}}$, which induce the emergence of the EP3.
Considering the weak-excitation limit and $|\psi_{\textrm{tot}} \rangle \simeq |\textrm{g}_{1}, 0 \rangle + \alpha |\textrm{g}_{2}, 0 \rangle + \beta |\textrm{e}, 0 \rangle  + \delta | \textrm{g}_{1}, 1 \rangle$, the bases that contribute to the dynamics of the system are

\begin{equation}
	|1\rangle = |\textrm{g}_{2}, 0 \rangle \quad |2\rangle = |\textrm{e}, 0 \rangle \quad |3\rangle = |\textrm{g}_{1}, 1 \rangle.
\end{equation}


The characteristic polynomial of $H_{\rm{nH}}$ is given by

\begin{equation}
	\lambda^{3} + i \left( \gamma + \kappa \right) \lambda^{2} - \left( \frac{\Omega^{2}}{4} + g^2 + \gamma\kappa \right) \lambda - \frac{i \Omega^2\kappa}{4} = 0 
	\label{eq:ch_po}
\end{equation}

In the EP3 condition, Eq.~\eqref{eq:ch_po} is identical with $(\lambda-\lambda_{\rm{EP3}})^3 = 0$. 
Comparing the coefficients of the terms $\lambda^{2}$ and last terms, we obtain


\begin{equation}
	\lambda_{\rm{EP3}} = -i\frac{\gamma + \kappa }{3}
	\quad 
	\textrm{and}
	\quad
	\Omega_{\rm{EP3}} = \frac{2(\gamma + \kappa )}{3} \sqrt{  \frac{ \gamma + \kappa } {3  \kappa  }}.
	\label{lambda_omega_EP}
\end{equation}

Inserting Eq.~\eqref{lambda_omega_EP} to Eq.~\eqref{eq:ch_po} and comparing the coefficient of the term $\lambda^{2}$ to that of $(\lambda-\lambda_{\rm{EP3}})^3 = 0$, we also find

\begin{equation}
	g_{\textrm{EP3}} = \sqrt{   \frac{(\gamma + \kappa )^2}{3}    -   \frac{ (\gamma + \kappa )^3}{27\kappa }      -  \gamma  \kappa  } .
\end{equation}

\subsection{Eigenvector}

Given $\lambda_{i}$, we represent the eigenvector $| \psi_{i} \rangle = C_{i} \left( c_{1,i}|1\rangle +  c_{2,i}|2\rangle + c_{3,i}|3 \rangle \right)$ with the normalization coefficient $C_{i}$.
The analytic form of $c_{1,i}, c_{2,i}, c_{3,i},$ and the decomposition matrix $\Xi$ are expressed as

\begin{equation}
	\begin{pmatrix} 
		c_{1,i} \\
		c_{2,i}  \\
		c_{3,i}
	\end{pmatrix} = 
	\begin{pmatrix} 
		\frac{\Omega}{2} \frac{\lambda_{i} + i\kappa }{ \lambda_{i} } \\
		\lambda_{i} + i\kappa \\
		g 
	\end{pmatrix}
	\quad\textrm{and}
	\quad
	\Xi=
	\begin{pmatrix} 
		\frac{\Omega}{2} \frac{\lambda_{1}+i\kappa }{\lambda_{1}} & \frac{\Omega}{2} \frac{\lambda_{2}+i \kappa }{\lambda_{2}} &  \frac{\Omega}{2} \frac{\lambda_{3} + i\kappa }{ \lambda_{3} }   \\
		\lambda_{1}+i\kappa & \lambda_{2}+i\kappa & \lambda_{3} + i \kappa  \\
		g & g & g  \\
	\end{pmatrix},
\end{equation}

\noindent
allowing us to construct the non-Hermitian Hamiltonian through $\Xi \cdot \Lambda \cdot \Xi^{-1} = H_{\rm{nH}}$, where

\begin{equation}
	\Lambda =
	\begin{pmatrix} 
		\lambda_{1} & 0 & 0  \\
		0 & \lambda_{2} & 0  \\
		0 & 0 & \lambda_{3}  \\
	\end{pmatrix}.
\end{equation}

Thus the coalesced eigenvector at the EP3 is
\begin{equation}
	\begin{pmatrix} 
		c_{1, \textrm{EP3}} \\
		c_{2, \textrm{EP3}}  \\
		c_{3, \textrm{EP3}}
	\end{pmatrix} = 
	\begin{pmatrix} 
		\frac{\Omega_{\textrm{EP3}}}{2} \frac{\lambda_{\textrm{EP3}} + i\kappa }{ \lambda_{\textrm{EP3}}}  \\
		\lambda_{\textrm{EP3}} + i\kappa \\
		g_{\textrm{EP3}} 
	\end{pmatrix},
\end{equation}

\noindent
with the omission of $C_{i}$.





\section{Quantum Monte Carlo Simulation}

In order to obtain the data in Fig.~3 of the main text, we perform quantum Monte Carlo (qMC) simulation with python QuTip. 
The employed Hamiltonian includes the weak cavity probing term in addition to Eq.~\eqref{eq:Htot},

\begin{equation}
	H_{\textrm{tot, qMC}} = \omega_{1} \sigma_{\textrm{e}} + (\omega_{1}-\omega_{2}) \sigma_{\textrm{g}_{2}} +  \omega_{\textrm{c}} a^{\dagger}a
	+ g ( a^{\dagger}\sigma_{\textrm{g}_{1}\textrm{e}} + \textrm{h.~c.} )
	+ \frac{\Omega}{2} ( \sigma_{\textrm{g}_{2}\textrm{e}} e^{i\omega_{\textrm{p}\rm{'}}t} + \textrm{h.~c.} )
	+ \epsilon  ( a e^{i \omega_{\textrm{p}} t} + \textrm{h.~c.} ),
	\label{eq:Htotqmc}
\end{equation}


\noindent
where $\epsilon$ is the amplitude of the probing field, $\omega_{\textrm{p}}$ is the frequency of the probe laser. 
Defining $U = \textrm{exp} [ i ( \omega_{\textrm{p}}\sigma_{\textrm{e}}$ \\
+ $(\omega_{\textrm{p}} - \omega_{\rm{p'}}) \sigma_{\textrm{g}_{2}} +  \omega_{\textrm{p}} a^{\dagger}a ) t]$, we obtain the interaction Hamiltonian

\begin{equation}
	H_{\textrm{int, qMC}} = -\Delta \left( \sigma_{ \textrm{e} }  +  \sigma_{ \textrm{g}_{2} }   +  a^{\dagger}a  \right) 
	+ g \left( a^{\dagger} \sigma_{\rm{g}_{1}\rm{e}} + \textrm{h.~c.} \right)
	+ \frac{\Omega}{2} \left( \sigma_{\textrm{g}_{2}\textrm{e}} + \textrm{h.~c.} \right)
	+\epsilon (a + \textrm{h.~c.}),
	\label{eq:Hintqmc}
\end{equation}


\noindent
with $\Delta = \omega_{\textrm{p}}-\omega_{\textrm{c}}$. 
Eq.~\eqref{eq:Hintqmc} is actually used in the qMC simulation.
The atomic and cavity decay operators are identical with those of Eq.~\eqref{eq:eom}.

\section{Fitting function of cavity-transmission spectrum}

We derive the fitting function for the cavity-transmission spectrum, Eq.~(3) of the main text, which is needed for extracting the eigenvalues. 
When the amplitude of the probe field is very weak such that the cavity mean photon number is much less than unity, the equation of motion for $|\psi \rangle =  \alpha |\textrm{g}_{2}, 0 \rangle + \beta | \textrm{e}, 0 \rangle + \delta | \textrm{g}_{1}, 1 \rangle$ can be expressed as

\begin{equation}
	\frac{d}{dt} |\psi\rangle = -i(H_{\rm{nH}} - \Delta I) |\psi\rangle -i
	\begin{pmatrix}
		0  \\
		0  \\
		\epsilon \\
	\end{pmatrix},
	\quad
	|\psi\rangle =	
	\begin{pmatrix}
		\alpha  \\
		\beta  \\
		\delta \\
	\end{pmatrix},
\end{equation}

\noindent
with the identity matrix $I$, and $\epsilon \ll \kappa$.

In the steady state, we find the state vector
\begin{equation}
	\label{eq:eos_trans}
	\begin{pmatrix}
		\alpha_{\textrm{ss}} \\
		\beta_{\textrm{ss}} \\
		\delta_{\textrm{ss}} \\
	\end{pmatrix}
	= (\Xi \Lambda \Xi^{-1} - \Delta I)^{-1}
	\begin{pmatrix}
		0 \\
		0 \\
		-\epsilon\\
	\end{pmatrix}
	=\Xi(\Lambda - \Delta I)^{-1} \Xi^{-1}
	\begin{pmatrix}
		0 \\
		0 \\
		-\epsilon\\
	\end{pmatrix},
\end{equation}

\noindent
where $\alpha_{\textrm{ss}}$, $\beta_{\textrm{ss}}$, and $\gamma_{\textrm{ss}}$ are the steady-state coefficients. 
From the analytic solution of Eq.~\eqref{eq:eos_trans}, we find

\begin{equation}
	\delta_{\textrm{ss}} = - \frac{\Delta ( i \kappa -\Delta + \lambda_{2} + \lambda_{3}  )\kappa + \lambda_{1} (\Delta \kappa -i \lambda_{2} \lambda_{3}  ) }{(\Delta - \lambda_{1})(\Delta - \lambda_{2}) (\Delta - \lambda_{3} ) \kappa} \epsilon.
\end{equation}

This leads to the cavity-transmission spectrum
\begin{equation}
	T = \left|     \frac{\delta_{\textrm{ss}}}{\epsilon / \kappa  }  \right|^{2} = \left | \frac{\Delta (  i\kappa    -\Delta + \lambda_{2} + \lambda_{3}  ) \kappa + \lambda_{1} ( \Delta \kappa -i \lambda_{2} \lambda_{3} ) }{(\Delta - \lambda_{1})(\Delta - \lambda_{2}) (\Delta - \lambda_{3} ) } \right |^{2},
	\label{eq:fitting_model}
\end{equation}

\noindent
with the eigenvalues of $\lambda_{1}$, $\lambda_{2}$, and $\lambda_{3}$. 
Via fitting the qMC simulation data with Eq.~\eqref{eq:fitting_model}, identical with Eq.~(3) of the main text, we obtain the three eigenvalues.

\section{Extracting eigenvalues with calcium ion}

\begin{figure*} [h]
	\includegraphics[width=6.3in]{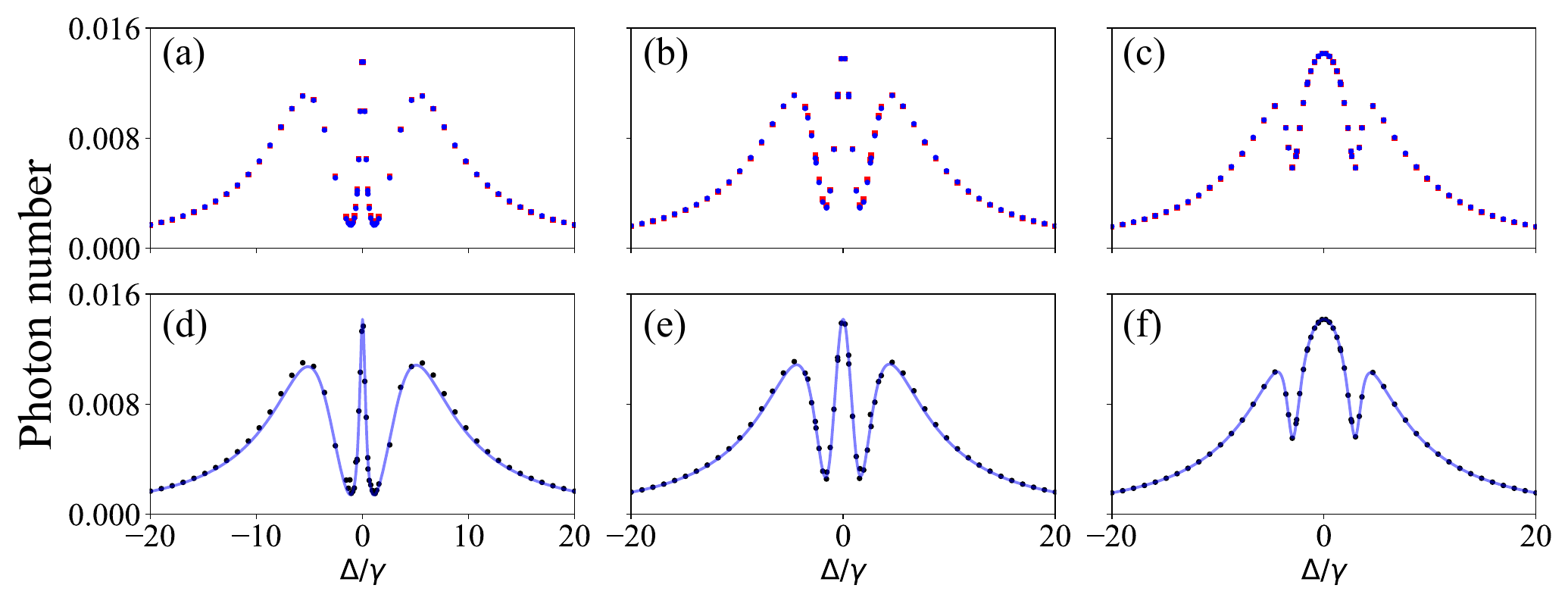} 
	\caption{
		(a)--(c) Cavity photon number (cavity transmission) as a function of $\Delta (=\omega_{\textrm{p}} - \omega_{\textrm{c}})$.
		Red points are results of numerical simulations with actual value of $\gamma_{2}$ of $^{40}$Ca$^{+}$, and blue points are with $\gamma_{2}=0$.
		(d)--(f) Quantum Monte Carlo simulation results with $\gamma_{2}$ are black points, blue line is fitting result with Eq.~\eqref{eq:fitting_model}.
		Parameters are $g/\gamma=4.60$ and $\Omega/\gamma=2.00$ in (a), (d), and $g_{\rm{EP3}}/\gamma=3.38$ and $\Omega_{\rm{EP3}}/\gamma=3.27$ in (b), (e), and $g/\gamma=2.00$ and $\Omega/\gamma=6.00$ in (c), (f).
		Spectrum of (b) and (e) are obtained at the EP3 condition.
		$\kappa/\gamma=6.94$. 
	}
	\label{fig:spectrum_ca}
\end{figure*}

\begin{figure} [h]
	\includegraphics[width=6.3in]{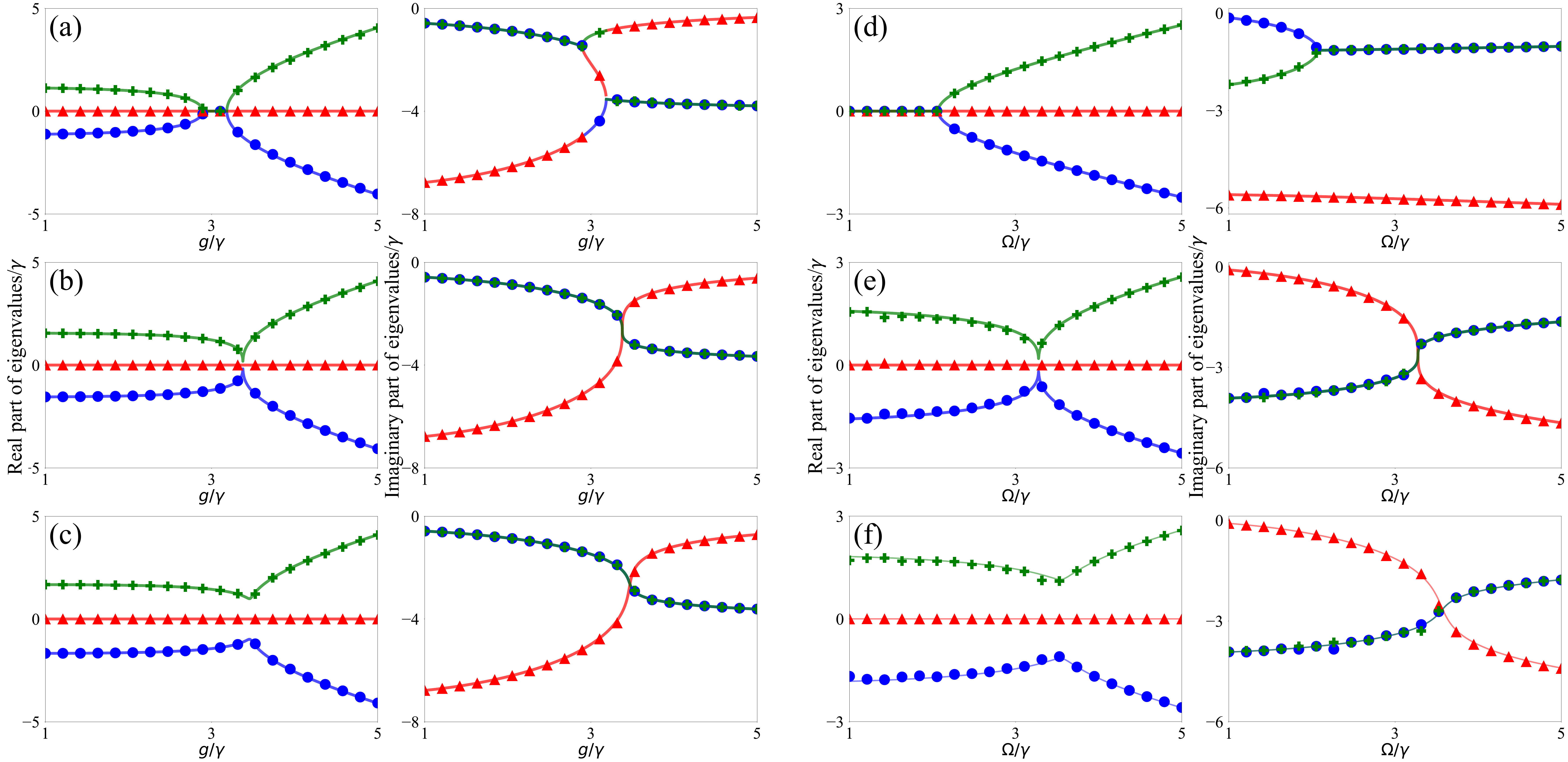} 
	\caption{
		Real and imaginary parts of the numerically obtained eigenvalues of $H_{\rm{nH}}$ are shown in lines.
		Green line is $\lambda_{1}$, red is $\lambda_{2}$, and blue is $\lambda_{3}$.
		Points are obtained from fitting the transmission spectrum, the result of quantum Monte Carlo simulation with actual values of $^{40}$Ca$^{+}$.  
		(a) $\Omega/\gamma = 2.50$, (b) $\Omega/\gamma= 3.27$, (c) $\Omega/\gamma=3.50$, (d) $g/\gamma = 2.50$, (e) $g/\gamma = 3.38$, and (f) $g/\gamma = 3.50$.
		EP3 appears in (b) and (e). 
		$\kappa/\gamma=6.94$. 
	}
	\label{fig:comparison_ca}
\end{figure}

In the qMC simulations of the main text, the decay rate from $|\textrm{e}\rangle$ to $|\textrm{g}_{2}\rangle$ $(\gamma_{2})$ is neglected.
In order to examine the influence of $\gamma_{2}$ to the cavity transmission spectrum, correspondingly to the extraction of the eigenvalues, we herein perform numerical simulations including the actual value of $\gamma_{2}$. 
All other parameters are identical with those in the main text. 
In Figs.~\ref{fig:spectrum_ca}(a)--(c), we obtain the cavity transmission spectrum with and without $\gamma_{2}$, compare those two results, and find that the difference is negligibly small. 
Therefore, this comparison confirms that the actual experiment with $^{40}$Ca$^{+}$ ion is suitable for extracting the eigenvalues of $H_{\textrm{nH}}$.
We fit the transmission spectrum of actual values of  $^{40}$Ca$^{+}$ ion with Eq.~\eqref{eq:fitting_model}, showing excellent agreement [Fig.~\ref{fig:spectrum_ca}(d)--(f)].
The extraction results of all eigenvalues for several $g$ and $\Omega$ are presented in Fig.~\ref{fig:comparison_ca}.

\end{document}